\documentclass[11pt]{article}
\usepackage{montmoriond,epsfig}

\def\be{\begin{equation}}
\def\ee{\end{equation}}
\def\bea{\begin{eqnarray}}
\def\eea{\end{eqnarray}}

\begin{document}
\vspace*{4cm}
\title{XXXVIIi\'eme Rencontres de Moriond: Electroweak 2002 Conference Summary}

\author{ HUGH E. MONTGOMERY}

\address{Fermi National Accelerator Laboratory, P.O.Box 500, Batavia, Il60510, U.S.A.
}

\maketitle\abstracts{ A substantial body of data is described
by the {\it Standard Model} of particle physics. However the
description is far from perfect and there is a growing number
of internal inconsistencies. These fall short of qualifying as
discoveries; nevertheless, examination of their merits is both
interesting and worthwhile. The existence of three families of quarks
and leptons is not understood. There are new data, especially from the
B factories; the latter are shining new light on the problem. From
several experiments, data show that our thoughts about the existence
of transitions between neutrino flavors, oscillations, may be correct
but the understanding of the patterns needs work. However, we see the
opening of a number of avenues of investigation as new facilities and
experiments come online. }

\section{Introduction}               

A conference summary, even in this case, in which the summary is
limited to the experimental talks and presentations, is
problematic. Neither the talk nor the written summary do justice to
the total content of the individual contributions.  Nevertheless, in
this paper, I attempt to give a sense of the physics I have learned
from the more than fifty talks about experiments presented at the
XXXVIIi\'eme Rencontres de Moriond (Electroweak).

The paper is organized such that I start with a brief discussion
of a number of {\it Searches for  New Phenomena} which,
unfortunately, did not find anything which goes beyond our current
description of the world. This is followed by a discussion of the
advances in {\it Neutrino Physics}. A sub-field enjoying focussed
attention is that of the {\it Quark Flavor Physics} of quarks. There are
beautiful measurements and some surprises as competitive
experiments present independent preliminary results and these are
discussed in Section 4. In section 5, we embark on a discussion of
the {\it Precision Electroweak Measurements}. It was a surprise to me that
it is difficult to make a clear distinction between the precision
measurements and what I call {\it Puzzles}. These measurements,
which do not fit well into the standard model, are discussed
in Section 6 followed by consideration of the way forward with
{\it New Starts} being discussed in Section 7. Finally, Section 8
contains a very brief {\it Perspective}.

\section{Searches for New Phenomena}

Among the searches~\cite{loidl} for weakly interacting massive
particles(WIMPs), the results from the DAMA~\cite{DAMA} experiment,
which initially indicated a possible signal, are now not generally
accepted by the community.  The CDMS experiment has
limits~\cite{CDMS}, which exclude much of the phase space favored by
DAMA, and at the time of writing, there are rumors of new results from
the EDELWEISS experiment. The regions of sensitivity are usually
displayed in the plane of cross section of the WIMP with matter and of
the mass of the WIMP. There are many experiments planned to extend
the reach in WIMP-nucleon cross section, from about $10^{-41}cm^{-2}$,
by several orders of magnitude. Prerequisites are low backgrounds; the
concurrent use of two detection techniques, to combat that background,
is generally expected. For example, some experiments use cryogenic germanium
detectors and detect both the recoil/phonon and the ionization
signals. The greatest sensitivity for all these experiments is in the
range of 20-90 $GeV$.

The couplings between the gauge bosons of the electroweak model
are completely prescribed by the theory. There have been studies
at the Tevatron Collider, which established~\cite{tgcreview}
the non-Abelian nature of the couplings. At LEP, the couplings are
directly measured~\cite{weber} and the central values correspond
well with expectations.

The Tevatron collider experiments have traditionally performed
especially sensitive searches for signals of new physics involving
strongly interacting partons, quarks and gluons. New
results~\cite{hagopian,velev} continue to consolidate these searches
using data from 1992-96, Run I of the Tevatron Collider.  The new
fashion is to search for extra dimensions through the actual
production of Kaluza-Klein gravitons in the final state.  These could
be in conjunction with either jets or vector bosons.  In both cases,
missing transverse energy is a key signature.  Limits of several
hundred GeV are obtained. These are similar to equivalent
limits~\cite{leplsg} from the LEP experiments.  Specifically, from LEP,
we see limits of about 800 GeV on the masses in low scale gravity
models. In all these searches, it has been demonstrated that the use
of angular distributions is a powerful tool with which to discriminate
against backgrounds.

 There are new results~\cite{boscherini} from the HERA experiments, H1
 and ZEUS, on the search for leptoquarks. This is the classic
 opportunity for this machine, which nicely complements the
 measurements at the Tevatron. There are also some new limits on the
 masses of excited leptons, both charged and neutral (heavy neutrinos)
 in the range 200-250 GeV for couplings of order $10^{-2}$ to
 $10^{-3}$ times the electroweak coupling.

\begin{figure}
\hskip 1.7in
\psfig{figure=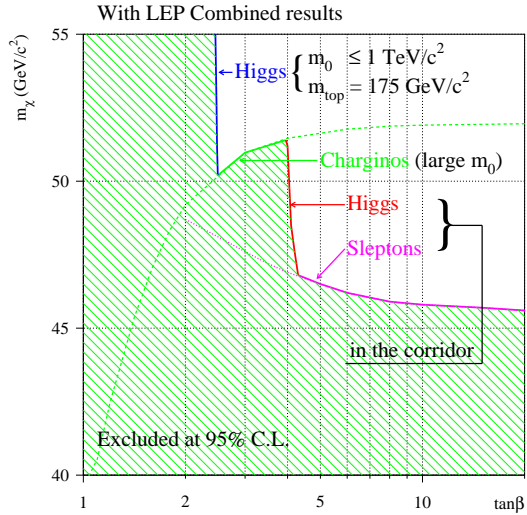,height=3in} 
\caption{Lightest Supersymmetric Particle(LSP) Combined Limit from LEP experiments.}\label{fig:leplsp}
\end{figure}

The parameter space associated with supersymmetric models is enormous,
even when rationalizations and approximations are made.  The
minimial supersymmetric standard model, in particular, the
"supergravity" species, requires the specification of five or so
parameters. Generally applicable representations of the results are
important and, with the now very mature analyses~\cite{latkineh}, we
see that the kinematic limits are being approached, almost independently
of the channel, for most scenarios. The lightest supersymmetric
particle is a key state in many models; it is also thought to be a
candidate for identification as the embodiment of cold dark matter. At
arbitrarily high values of $tan~{\beta}$, the ratio of the vacuum
expectation values associated with the two Higgs doublets in this
model, and thus essentially independent of other model parameter
values, the lower mass limit is 56 GeV at $95\%$ confidence level.
The experimental mass limit, as a function of $tan~{\beta}$, is shown
in Fig.~\ref{fig:leplsp}.

What distinguishes one supersymmetric model from another for the
theorist is the mechanism by which the symmetry is broken. A popular
class of models falls under the designation ``gauge mediated
supersymmetry breaking". These models tend to lead to a preponderance
of photons in the final state, but they can also lead to relatively
long lifetimes for some of the sparticles. The extra parameter means
that the experimental groups present the results~\cite{katyaklein} of
their searches in terms of cross section limits as a function of both
mass and lifetime. Typical lower limits for the mass of the sleptons
in the range of 60-90 GeV have been obtained.

 Initially the searches for supersymmetric particles concentrated on
 the pair production of supersymmetric partners of which the decays
 were constrained to conserve R-parity($R_p$), the quantum number,
 which distinguishes a particle from a super-partner. The corollary is
 that the cascade decays should terminate with a neutral
 lightest-supersymmetric particle(LSP), of relatively low mass, which
 escapes detection and thus leaves missing transverse energy as its
 signature. Admitting the non-conservation of R-parity deprives the
 experimentalist of this rather distinctive signature.  Nevertheless,
 the increasing sophistication of searches for supersymmetric
 particles is demonstrated by the fact that the current LEP
 limits~\cite{fouchez} on the masses of sparticles in $R_p$ violating
 models approach very closely the equivalent limits for the $R_p$
 conserving cases.

 The adjective "exotic" is often applied to those searches for which
 no clear, currently fashionable, theoretical justification can be
 found. However, that does not invalidate them; indeed for many people
 such possibilities are the most exciting. The searches at LEP yield
 80-100 GeV lower limits~\cite{giacomelli} on the masses of excited
 leptons, leptoquarks (cf. the searches from HERA and the Tevatron
 discussed above), heavy leptons, and technicolor. It is interesting
 to look for short-hand representations of negative results and in
 this spirit we can surmise that, lower mass limits ranging from 80
 GeV to 200 GeV have been set with the respective couplings varying
 from $10^{-3}$ to $1$.

 The multiplicity of models for the physics beyond the standard model
 have, as a corollary, just as many different scenarios for the
 structure in the Higgs sector. In the realm of SUSY, the parameters
 such as $tan~{\beta}$ can lead to different branching ratios. The
 Higgs particles may decay dominantly to bosons, or to leptons, to the
 first, second, or third generations. The presentation~\cite{quadt} of
 the results can be correspondingly complicated. However, it is
 reassuring to see that the generic models with one and two Higgs
 doublets, and searches with decay-mode-independent techniques, as well
 as flavor-independent techniques complement the specialised, Minimal
 Supersymmetric Standard Model and other, model dependent
 searches. The lower mass limits range from about 80 GeV up to the limit on
 the mass of the standard model Higgs at slightly greater than 114
 GeV.

\section{Neutrino Physics}

In preparing for this conference I looked back at a transparency
that I had used in 1999. At that time the neutrino oscillation
scene included three regions with possible positive signals. The
"LSND anomaly", is a putative signal, with $\Delta{m^2}$ greater
than $0.1 (eV)^2$. There were first indications of oscillations
with $\Delta{m^2}$ of a few $\times 10^{-3}~(eV)^2$ and a large mixing angle,
primarily from SuperKamiokande but with supporting results from
other experiments. Finally, the long standing solar
neutrino deficit suggested neutrino mixing at with a very small
mass difference squared: less than $10^{-4} (eV)^2$.

\begin{figure}
\hskip 1.0in
\psfig{figure=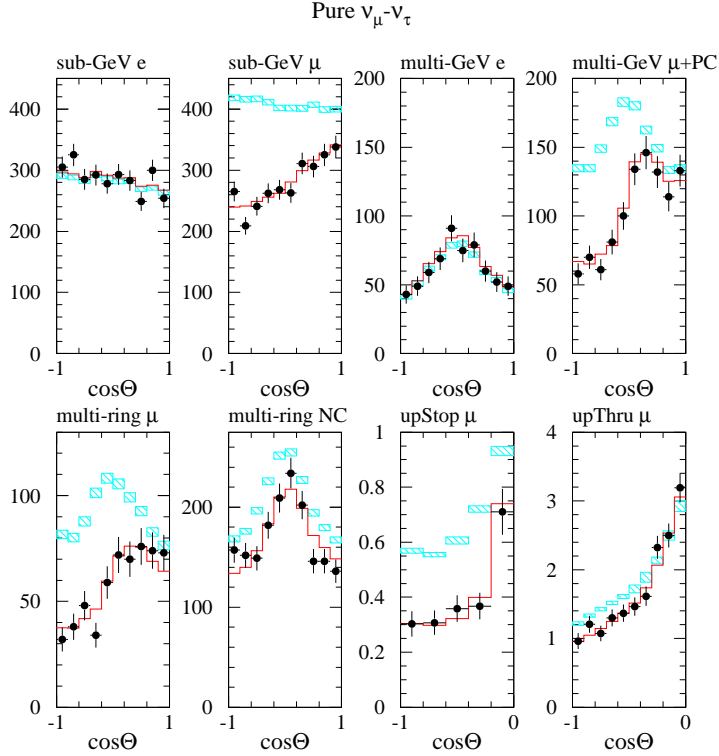,height=4in}
\caption{Data from the SuperKamiokande Experiment showing the agreement between several atmospheric data sets as a function of angle and a model of oscillations of $\nu_{\mu}$ into  $\nu_{\tau}$.}
\label{fig:superkatmos}
\end{figure}

At this conference the conclusions of the Chorus and Nomad
experiments at CERN were described~\cite{zuber}. These experiments
were designed with tau neutrino appearance in mind and they have
provided limits on the oscillations of both muon neutrinos, which
was the primary aim, and electron neutrinos, to tau neutrinos. For
mass difference squared above $40~eV^2$, they exclude, with $90\%$
confidence level, $\sin^2 (2\theta)> 5\times 10^{-4}$. For
electron neutrinos to oscillate to tau neutrinos the NOMAD limits,
which are the better of the two, are nearly an order of magnitude
less restrictive in each of the mass difference squared and the
square of twice the mixing angle. Finally the NOMAD limits for
muon neutrinos to oscillate to electron neutrinos exclude a
substantial fraction of the space occupied by the LSND results.

In the atmospheric neutrino arena, the array of SuperKamiokande
work~\cite{smy,aweber} was rapidly advancing until the recent problems
with the detector brought progress to a halt. These results include
not only the rates, but several measurements of the dependence of the
fluxes on angle or time. The results in parameter space are now really
beginning to look like a rather clear indication that
$\Delta{m^2}\simeq 2-3 \times 10^{-3}~(eV)^2$ and that $\sin^2
(2\theta)$ is close to unity. Several of their data sets are displayed
in Fig.~\ref{fig:superkatmos}. There is good agreement between the
data and the description provided by a model in which the muon
neutrino is transformed into the tau neutrino.

K2K is the first accelerator-based neutrino oscillation experiment
with a moderately long baseline. The experiment~\cite{ichikawa}
operates with approximately one GeV neutrinos. Fifty-six events are observed
in the far detector when eighty-one are expected. The probability for a null
oscillation hypothesis is 3\%; the result would be well described with
$\Delta{m^2}\simeq 3 \times 10^{-3}~(eV)^2$.  It seems that there is a
good plan to resume operation of SuperKamiokande in a year or so with
a reduced complement of photomultiplier tubes. Since the present
results seem to show the biggest effect at an energy of 0.6 GeV, K2K
also will rebuild their near detector to better match this energy.


The SuperKamiokande experiment has also produced prominent results
relevant to the understanding of the solar neutrino deficit. They
have measured angular distributions as well as making rate
comparisons. The SNO experiment~\cite{mcgregor} produced its first
results a year ago based on their measurement of charged current
interactions, which are sensitive only to electron neutrinos, and
elastic scattering. The latter is sensitive primarily to
electron neutrinos but also, with a relative sensitivity of about
15\%, to muon and tau neutrinos. In SNO the statistics for the
elastic scattering are rather limited. However, SuperKamiokande has
made a precise measurement of that quantity. Together the charged
current and elastic scattering measurements permit a solution for
both electron and non-electron neutrino fluxes. The results
beautifully demonstrate that the non-electron neutrino flux is
indeed non-zero. Further, when added to the electron neutrino flux,
the sum matches well to the standard solar model. In this way, a
convincing case is constructed, that neutrino flavor
transitions have been observed. Since the time of the conference,
SNO has presented~\cite{snonew} results on the measurement of
the neutral current neutrino interactions. These confirm and take
a step further than the earlier measurements.

\section{Quark Flavor Physics}

K mesons were discovered some fifty years ago. Their importance was
enhanced by the discovery of CP violation in the neutral kaon system
in 1964. With the discovery of the Upsilon, the $b$ quark joined the
game, and the physics of flavor started to be described by the, now
very famous, Cabibbo-Kobayashi-Maskawa matrix, which relates the strong
interaction, mass, eigenstates to the weak interaction states of the
three families.

\begin{figure}
\hskip 1.5in
\psfig{figure=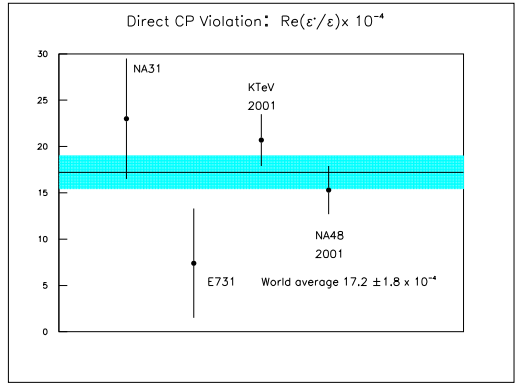,height=2.0in}
\caption{Recent Results on
$\cal{R}({\epsilon}${'}/$\epsilon$).}\label{fig:kaonepsprime}
\end{figure}

The recent observation~\cite{bnle787} by BNL-E787 of a second
example of the decay $K^{+}\rightarrow \pi^{+}\nu\bar\nu$ has been
interpreted by many as a demonstration that a substantive
measurement of that decay rate is just around the corner; a bright
future is presaged for this avenue of investigation, perhaps with
a next-generation experiment, dubbed CKM, at Fermilab. Such an
experiment~\cite{dukes} is thought to provide a theoretically clean
measurement of one of the sides of ``the" unitarity triangle
representing the CKM matrix.

There have been theoretical predictions suggesting that an accessible
degree of CP violation might occur in hyperon decays.  An experiment
called Hyper-CP has been performed at Fermilab, which has
measured~\cite{dukes} the branching fraction for the flavor changing
neutral current decay $K^{+} \rightarrow
\pi^{+}\mu^{+}\mu^{-}$. The result, $(9.8 \pm 1.0(stat) \pm
0.5(syst))\times 10^{-8}$ appears to distinguish between two
previous measurements, which were at variance.

Another initiative in search of CP violation is the KLOE experiment at
the DA$\Phi$NE accelerator. While accumulating the requisite
luminosity for the CP violation measurement, this experiment, which
will exploit the mutual tagging of neutral kaons from the decay of the
$\phi$ particle, has measured~\cite{patera} a number of radiative
$\phi$ decays as well as the rare kaon branching fraction,
BR($K_{s}\rightarrow
\pi{e}\nu) = (6.79\pm 0.33 (stat) \pm 0.16 (syst))\times
10^{-4}$.

There are two recent experiments, which have measured CP violation in
the neutral $K_{L}$ system. The NA48 at CERN has
results~\cite{graziani}, which cover their data through 1999 but which
do not incorporate any information from their running in 2001 with a
reconstituted spectrometer, after their accident of 2000. The KTeV
experiment at Fermilab has results~\cite{whitmore}, which are limited
to their data through 1997; there are more data taken through
1999. These are the latest, and perhaps last, in a series of
experiments at the two laboratories, which have continued over 20
years. Agreement between the two has not always been evident, but the
current results, illustrated in Fig.~\ref{fig:kaonepsprime}, are in
accord at a confidence level of 13{\%}, and yield a world average of
$\cal{R}({\epsilon}${'}/$\epsilon$) = $(17.2 \pm 1.8) \times 10^{-4}$.


The two experiments also contribute numerous other measurements of the
neutral kaon system. For example, they have measured a large
number of branching fractions of rare decays. The decay $K_L
\rightarrow \mu\mu$ depends on different components, one of which,
in turn, depends on the $K{{\gamma}^{*}}{{\gamma}^{*}}$ vertex.
This vertex is determined by a measurement of the branching
fraction, $BF(K^{0}_{L} \rightarrow ee\mu\mu) = (2.62 \pm
0.23(stat) \pm 0.18(syst)) \times 10^{-9}$. In addition, there are
vital parameters, such as the mass differences and lifetimes associated
with the different states $K_{L}$ and $K_{S}$, which are now known
with exquisite precision.

Both the KEK-B and PEP-II asymmetric electron-positron colliders
are working extremely well. Almost step for step, the luminosity of
each, and the integrated luminosity analysed by the two
experiments, BELLE and BaBar respectively, has increased. It is an
impressive performance by all.

\begin{figure}
\hskip 1.7in
\psfig{figure=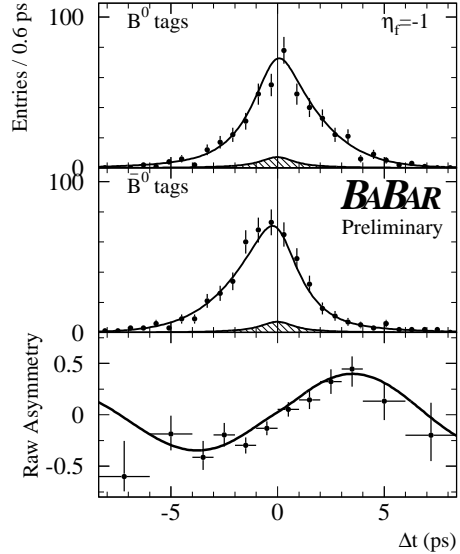,height=3in}
\caption{The CP asymmetry for $B\rightarrow (J/\psi, \psi(2S), \chi_{c1})K^{0}_S$ decays, as observed by the BaBar experiment.}
\label{fig:bfacone}
\end{figure}

As anticipated the first results in CP violation appeared in the
classic $B \rightarrow J/\psi~K_{S}$ channel, which determines the
parameter $sin~2\beta$ ( When in Rome ....; I will follow the
European/North American convention). While a year ago, there were
differences between the two experiments, at this conference BELLE
reported~\cite{trabelsi} $sin~2\beta = 0.82 \pm 0.12(stat) \pm 0.05
(syst)$ while BaBar reported~\cite{raven} $sin~2\beta = 0.75 \pm
0.09(stat) \pm 0.04(syst)$. The quality of the BaBar results is
illustrated in Fig.~\ref{fig:bfacone}. The agreement is now
satisfactory.  The changes to the results from each experiment, came
not only from the increased statistics accrued during the past year
but also from refinements to the analyses.

\begin{figure}
\hskip 1in
\psfig{figure=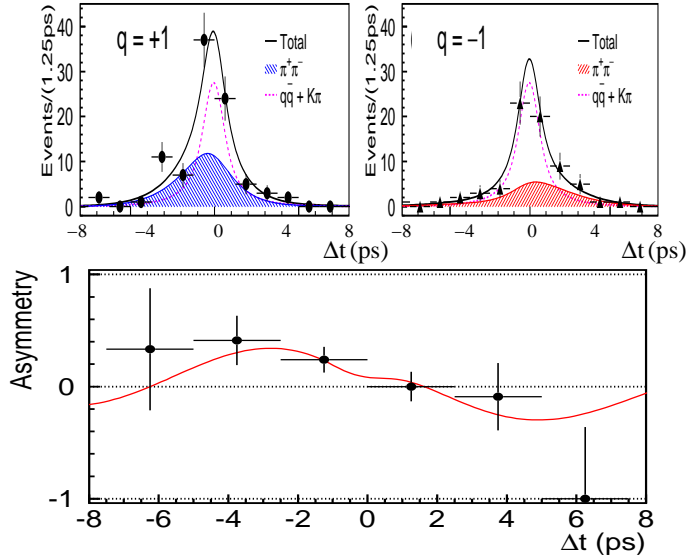,height=3in}
\caption{Measured CP Asymmetry in the Decay $B\rightarrow \pi^{+}\pi^{- }$ as a function of the time difference as measured by the BELLE experiment.} \label{fig:bfactwo}
\end{figure}

The decay $B~\rightarrow~\pi\pi$ is related to the CKM parameter
$sin~2\alpha$; unfortunately this relationship is not as clean as
that for $sin~2\beta$. There are several different phases at work
since the loop (penguin) diagrams are expected to play a large
r\^ole. The data are fit for both sine and cosine variation; the
latter indicates the possible direct CP violation component. BaBar
sees~\cite{farbin} essentially no CP violation with the results $S=
-0.01 \pm 0.37(stat) \pm 0.07(syst)$ and $C = -0.02 \pm 0.29(stat)
\pm 0.07(syst)$. In contrast BELLE sees substantial CP violation,
indeed, their fit results~\cite{ishikawa} violate a constraint that
the quadratic sum of C and S not exceed unity. They find $S = -1.21
+0.38 - 0.27(stat) +0.16 -0.13(syst)$ and $ C = -0.94 \pm +0.25
-0.31(stat) \pm 0.09(syst) $ (Again we use the sign convention used by
BaBar, which changes the sign of the cosine coefficient presented by
Belle). The results from BELLE are shown in Fig.~\ref{fig:bfactwo}.

The B factory experiments and their predecessor CLEO have a broad
physics program. The yield of final states with charm is
impressive~\cite{pompili} and the resulting physics is very
interesting~\cite{ligeti}. In this context the recent B-factory
measurements of $D^{0}-\bar{D^{0}}$ mixing now have uncertainties of about
$\pm~1\%$, slightly less than the previous Fermilab experiments, FOCUS
and E791(FNAL), and of CLEO. Thus far their results~\cite{pompili} are
consistent with zero at this level.

Beyond the reach of the B factories are the $B_s$ states and the
mixing between the neutral $B_s$ mesons, which is controlled by the
length of one side of the unitarity triangle. Results come from
LEP~\cite{sciaba}, and from the SLD~\cite{chou} experiment at
SLAC. The results from the SLD experiment are not quite final; the
present limit, based on the world data, gives a limit of
$\Delta{m_s}~>~14.9~ps^{-1}$ at $95\%$ confidence level, on the mixing
parameter.

The study of the CKM matrix, will eventually depend on the
interweaving of many different measurements. In addition to the
much touted measurements of  $B \rightarrow J/\psi K_{S}$ or $B
\rightarrow \pi\pi$, many of the rare decays can shed light on
different aspects of the problem. By using more than one
measurement, the relevant CP violating parameters are indirectly
accessible. This has led to a recognition that a systematic
approach, the {\it CKM Fitter}~\cite{laplace} is needed. In
particular this initiative leads immediately to the incorporation
of many of the rare B decay
measurements~\cite{tanaka,huang,schwartoff} from BaBar, BELLE, and
CLEO as well as rare K decays.

 Meanwhile the measurements involving electroweak penguins,
for example those of $b \rightarrow s\gamma$ transitions,
immediately provide search windows beyond the standard model. We
would expect the particles associated with the new physics to
participate in the loops. Thus far nothing unexpected has been
been seen~\cite{tanaka,huang}.

\section{Precision Electroweak Measurements}

An important barometer of our understanding and comfort with the
standard model is provided by the fit of a broad spectrum of
electroweak data by the LEP Electro-Weak Working Group(EWWG).

The recently updated fit~\cite{myatt} contains several new components;
however, let us begin with a well established parameter, the
electromagnetic coupling constant $\alpha_{em}(m_Z)$ evaluated at the
mass of the $Z$. $\alpha_{em}$ is well measured at low momentum
transfers, however the extrapolation to $m_Z$ involves a dispersion
integral, which is usually evaluated using input from the experimental
measurement of the ratio R of hadron production and muon pair
production in electron-positron collisions. A similar, but not
identical, integral also enters  the hadronic component of the
anomalous magnetic moment of the muon, $(g-2)_{\mu}$. The precision of
the extrapolation has been improved~\cite{hu} by the recent high
precision data from the BES experiment on the BEPC collider in
Beijing. These data are in the centre of mass energy range from 2 to 5
GeV. With uncertainties of about 6\% on each measurement point, the
contribution to the integral from this energy range has been reduced
from more than 50\% to less than 30\% of the total error. This has
prompted some imaginative attempts to obtain good measurements in
other energy ranges. There are attempts to use initial state radiation
in both the KLOE experiment~\cite{valeriano} and the BaBar
experiment~\cite{buchmuller}.  The studies are in the preliminary
stages, indeed at this conference there was considerable discussion as
to how well the initial state radiation can be understood. Is the
uncertainty near 2-3 \% or much smaller? The energy ranges targeted
are from threshold for $\pi\pi$ production to 1 GeV at KLOE, and in
the ranges 1-2 GeV and 2-6 GeV at BaBar.

New this year are the direct measurements~\cite{parkes} of the $W$
boson width from LEP and we can expect that the equivalent
measurements from the Tevatron will also be incorporated soon. The $W$
masses themselves are not yet final~\cite{parkes}. There remain some
final evaluations of systematic uncertainties by the experiments,
which are expected soon. Of some concern are the possible effects of
Bose-Einstein correlations and color recombination.  In the former
case, the effects are observed~\cite{dierckxsens} when the two
particles originate from a single $W$ decay, but not when the the two
pions originate from different $W$ bosons. Since I had understood
Bose-Einstein correlations to be simply related to the source size,
this appears bizarre to me. As for the attempts to determine the
color-recombination effects, the experiments appear to have a broad
range of results. Fortunately, even if it is decided that a strategic
retreat is necessary, relinquishing the all-hadronic decays would only
lead to an increase of a few MeV in the final LEP $W$ mass
uncertainty.

Also fresh are the determinations of the forward-backward asymmetries
$A^{0,b}_{fb}$ and $A^{0,c}_{fb}$, as a result of a new
analysis~\cite{ciulli,elsing} from ALEPH. In truth, the numbers have
changed very little as a result of this reanalysis.  Nevertheless, it
is clear that a vital piece of the story is the judgement as to which
errors are common to the several experiments and which are not. Those
in the former category do not decrease with multiple measurements. As
presented, the data sets seem to be remarkably consistent across the
LEP experiments; there is nothing remarkable about the ALEPH result.

\begin{figure}
\hskip 1in\psfig{figure=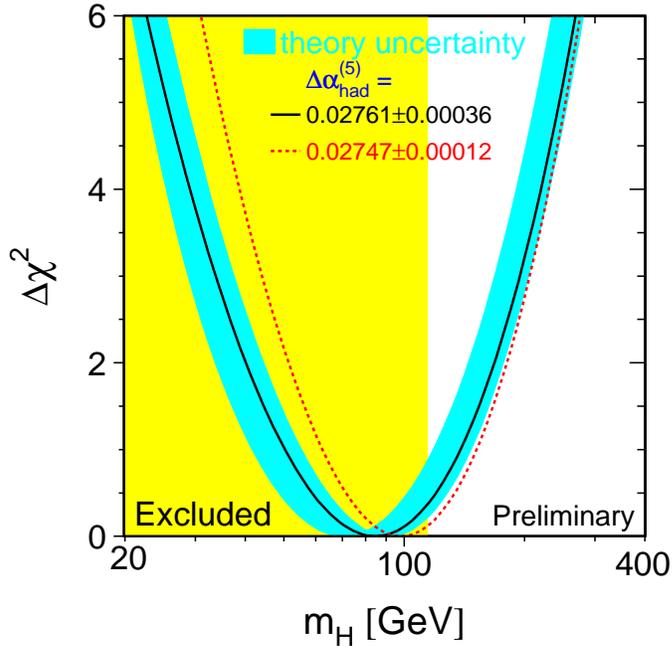 ,height=4in}
\caption{The Winter 2002, ElectroWeak Working Group fit showing
the $\Delta{\chi}^{2}$ as a function of Higgs mass, and the region of Higgs mass excluded by the direct searches.}
\label{fig:ewwg2002blueband}
\end{figure}

In the past year the neutrino scattering experiment, NuTeV at the
Fermilab Tevatron has presented new results~\cite{zeller} on the
determination of $sin^{2}\theta_{W}^{on-shell}$ = $0.2277 \pm
0.0013(stat) \pm 0.0009(syst)$, the electro-weak mixing angle. The
basic measurement is that of the ratio between the neutral and charged
current interactions using beams of both neutrinos and
antineutrinos. If expressed in terms of the $W$ mass, to which it is
closely related, the precision is of order 100 MeV, comparable to that
from any individual measurement from either LEP or the Tevatron
collider.

Finally there is a new value~\cite{myatt} for the atomic parity
violating parameter $Q_W$ measured with Cesium, which is also
incorporated in the fit.

The result of the global fit of all the 20 parameters (at the
highest level, many more when individual measurements are counted)
gives a ${\chi}^{2}$ of 29 for 15 degrees of freedom. This is a
poor fit! Although one can read off the curve of
$\Delta{\chi}^{2}$ that the mass of the Higgs should be less than
196 GeV with 95\% confidence level, it is my opinion that it would
be inappropriate to use such a fit to deduce such a limit. In
olden times, we used to scale the individual uncertainties by a
factor $\sqrt{\chi^{2}/n_{dof}}$ before making predictions!

Further, in an earlier section, the limit on the mass of a
standard model Higgs from a direct searches was given~\cite{quadt}
as about 114 GeV.   As shown in Fig.~\ref{fig:ewwg2002blueband}
this limit is higher than the Higgs mass which gives the best fit
of the standard model to the data.

\section{Puzzles}

\begin{figure}
\hskip 2.0in\psfig{figure=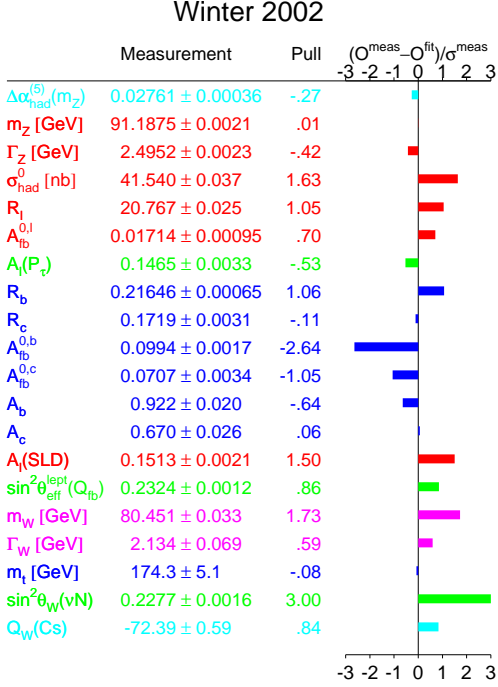 ,height=4in}
\caption{The Winter 2002, ElectroWeak Working Group fit showing
the pulls of the twenty experimentally determined parameters.}
\label{fig:ewwg2002pulls}
\end{figure}

Clearly the description of the previous section yields a puzzle.  Why
is the electroweak fit so bad? In straightforward terms the fit is bad
because, as shown in Fig.~\ref{fig:ewwg2002pulls}, the NuTeV
measurement, the forward-backward assymmetry for b quarks, and the $W$
mass differ from the central value by +3.0, -2.64, and +1.73 $\sigma$,
respectively. Chanowitz~\cite{chanowitz} has argued that the distinct
differences, the inconsistency, between the leptonic and heavy quark
determinations of the weak mixing, makes a physics statement. He also
points out that, were the heavy quark measurements to be rejected for
some reason, the central value of the Higgs mass would be much
lower. I am hesitant to follow this path because there are many
potential fault lines along which we might split the measurements;
and, however well motivated, this is only one of the
possibilities. Nevertheless, I would like to repeat my earlier
conclusion that the fit is currently so bad that I would not use the
implicit Higgs-mass limits for anything serious.

It is interesting that we are able to improve on classic
experiments. In the late 1950s, the observation of parity
violation in nuclear $\beta$ decay changed the thinking about
physics. A modern version of that same experiment, in which beams
of neutrons are allowed to decay in a magnetic field,
reports~\cite{abele} that its results are at variance with what one
deduces based on measurements with the 2nd and 3rd generations of
quarks. The experiment, in combination with the neutron lifetime
determines $V_{ud}$ = $0.9713 \pm  0.0013$. The expectation that
$V_{ud}^{2} +  V_{us}^{2} + V_{ub}^{2}$ =1, if the CKM matrix is
unitary, appears to be violated at the level of more than 2$\sigma$.

In the high energy regime, the Tevatron Collider experiments
continue to analyse their 1992-96 data and CDF has
found~\cite{velev} a couple of issues in their data. One is that in
the events with a $W$ boson and several jets, it is very difficult
to explain the yield of jets in which there are two $B$-tags of
the jet. The two tags are provided by the presence of a soft
lepton and by a large impact parameter. The probability of a
standard explanation is about 0.4\%. Similarly they have an
abnormal yield of events with both leptons, a photon, and missing
transverse energy. The anomaly has a probability of
0.7\% to be explained by standard sources.

\section{New Starts}

If we had no experiments in view, the puzzles described in the
previous section would truly represent a conundrum. Fortunately we are
in a position to discuss a number of initiatives which can hope to
illuminate our issues and to extend our horizons.

Over the past few  years, the HERA machine has been upgraded and
the experiments are poised~\cite{schultzcoulon} for a run which,
through 2006, would take them into the realm of large statistics
charged current measurements, and, using polarization, the
decomposition of the observed cross sections into the multiple
structure functions. The B factories have demonstrated superiority in
$e^{+}e^{-}$ collisions at the energy of the $\Upsilon(4S)$ so the
CESR accelerator and the CLEO experiment have refocussed
themselves~\cite{benslama} on the charm system and on a concerted
attack on numerous outstanding issues concerning the QCD description
of the bound states of quarks and gluons.

The neutrino field is rich in new experiments. The SNO experiment is
ongoing~\cite{mcgregor}, and, as we remarked earlier, is in a very
productive phase. The KamLAND experiment~\cite{dazely} starts to use
the neutrino fluxes from multiple reactors, and a new detector at the
Kamioke mine to examine, in the laboratory, the low mass solar
neutrino range of oscillations. Its sensitivity covers that of the
popular Large Mixing Angle(LMA) solution, in which $\Delta{m^2}$ is a
few $10^{-6}~(eV)^2$. The MiniBooNE detector~\cite{sorel} is taking
its first beam this spring as it gears up to resolve the LSND
mystery. The minimalists will hope for a negative result while others
will hope that the results demand a fourth neutrino-like state. The
definitive measurements concerning the intermediate mass mixing
solutions, used to describe the atmospheric neutrino phenomena may
well come from the new long-baseline accelerator-based
experiments~\cite{aweber} NUMI/MINOS(Fermilab to Soudan) and CNGS(CERN
to Gran Sasso) with the Icarus\cite{rico} and Opera experiments; the
former will be operating in about 3 years.  However, we should note
with pleasure that the Super-Kamiokande and K2K experiments will also
be operational by then.  Last-but-very-importantly, we are seeing
serious initiatives, the HARP experiment~\cite{radicioni} at CERN and
the MIPP-E907 experiment at Fermilab, to measure the relevant hadron
production cross sections and characteristics needed to determine the
all-important source terms for many of the neutrino oscillation
experiments.

The program of neutrino measurements, described above, can lead to an
understanding of the systematics of the neutrino sector.  Oscillations
imply mass and the determining parameter is a mass difference
squared. In order to obtain information on the masses themselves, we
must look elsewhere. The NEMO 3 experiment~\cite{piquemal} in the
Frejus Laboratory, not to be confused with the putative underwater
experiment in the Mediterranean Sea, is just starting to look for
neutrino-less double-$\beta$ decay. It incorporates important
ancillary measurements designed to ensure that all the unknowns are
systematically explored.

The energy frontier continues to attract a large fraction of our
field; for the present and for several years to come, that frontier is
at the Tevatron Collider. After an extensive upgrade to the
accelerator complex at Fermilab, Tevatron operations have been
reestablished during the past year. Although, as yet, the luminosity
is modest, the experiments, also extensively upgraded, are
operational. CDF shows off~\cite{rescigno}, among a number of
interesting distributions, clear charm signals produced online on the
basis of its new displaced vertex trigger. D{\O}
touts~\cite{verzocchi} mass distributions obtained using its new
tracking systems, silicon and scintillating fiber and, of course, its
new solenoid. Both experiments have the traditional high energy
signals of $W$ and $Z$ production and, with the jet signals at high
$E_T$, it seems that the benefit to high mass production, of the
modest(9\%) increase in the Tevatron energy is visible.

All these new experimental efforts generate a clear and justified
sense of expectancy in the field.

\section{Perspective}

When I accepted the task of giving the summary talk for this
conference, I anticipated that the challenge would be to find any
data, which could be taken as disagreeing slightly with our
standard model. What I found was different.

The neutrino field is clearly pushing to extend and enrich the
lepton sector. Incorporation of a neutrino mixing matrix into our
accepted paradigm, seems only to be a matter of time.

The advance of the B factory measurements is truly impressive; as is
often the case in our field, hero status goes to the accelerators as
much as to the experiments. Of course the youth of the experiments,
and of the analyses, ensures that agreement between the two
experiments does not always come with the first presentations; this is
healthy.

However, in the sectors already incorporated in a non-trivial
manner in the model, the description of the data, as it presently
stands, is distinctly ragged. The electroweak fit has a very low
probability, P($\chi^{2}$). In the section entitled {\it
Puzzles}, we enumerated the problems and also pointed out some
others which, for me, are new, such as the neutron $\beta$ decay
asymmetry result.

Fortunately, at the same time, we can see that we have a number of
experimental initiatives. In neutrino physics we have the
opportunity to immediately push further. In high energy hadron
collisions we are remounting an attack at the energy frontier
which has been dormant for five years. In the lower energy arena,
we are again demonstrating the ability of the field to maximize
the exploitation of its investments, no matter how venerable.

In all, I learned a lot and I also learned that the field is
healthy, where healthy means that it is not so easy to describe
all the data to hand, and we will soon have more such data.

\section*{Acknowledgments}

I was helped in the preparation of the talk by almost every one of the
speakers who gave an experimental talk and sometimes this help came in
advance from other individual members of the experiments.  In
addition, some of those who gave theory talks also helped my
education. It would be churlish to try and discriminate between all
these efforts; the help was generously given and is much
appreciated. I would like to thank my colleague Gene Fisk for
critically reading the manuscript.

This was my first ``Moriond" and I would like to thank the
organizers for their hospitality, the atmosphere was very
conducive to the free exchange of information and ideas.

This work was supported by the U.S. Department of Energy through Fermi
National Accelerator laboratory which is operated by Universities
Research Association Inc. under Contract No. DE-AC02-76CH03000 with
the United States Department of Energy.

\end{document}